\documentclass[a4paper,fleqn]{article} 
\usepackage{modsim}
\usepackage{times}
\usepackage{natbib} 
\usepackage{amsmath, amssymb, amsthm} 
\usepackage{bm}

\MODSIMhead{X. Luo and P.V. Shevchenko, Variable Annuity with GMWB: surrender or not, that is the question} 

\usepackage{rotating}
\usepackage{amsbsy,enumerate}
\usepackage{graphicx}
\usepackage{ccaption}
\usepackage{multirow}

\makeatletter
\renewcommand{\fnum@figure}[1]{\textbf{\figurename~\thefigure}. }
\renewcommand{\fnum@table}[1]{\textbf{\tablename~\thetable}. }
\makeatother

\begin{document}

\title{Variable Annuity with GMWB: surrender or not, that is the question}

\author{\underline{X. Luo} \address[A1]{{\it CSIRO Risk Analytics Group, Australia} \\This version: 31 July, 2015} and P.V. Shevchenko \addressmark[A1]} 

\email{Xiaolin.Luo@csiro.au} 

\date{July 2015}

\begin{keyword}
Variable annuity,
optimal stochastic control, optimal stopping time, bang-bang control, Guaranteed Minimum Withdrawal Benefit, surrender option,
Gauss-Hermite quadrature,  cubic spline.
\end{keyword}

\begin{abstract}
A variable annuity contract  with Guaranteed Minimum Withdrawal
Benefit (GMWB) promises to return the entire initial investment
through cash withdrawals during the policy life plus the remaining
account balance at maturity,  regardless of the portfolio
performance. We assume that market is complete in financial risk and
also there is no mortality risk (in the event of policyholder death,
the contract is maintained by beneficiary), thus the annuity price
can be expressed as an appropriate expectation. Under the optimal
withdrawal strategy of a policyholder, the pricing of variable
annuities with GMWB is an optimal stochastic control problem. The
surrender feature available in marketed products allows termination of the  contract before
maturity, making it also an optimal stopping problem.

Although the surrender feature is quite common in  variable annuity
contracts, there appears to be no published analysis and results for
this feature in GMWB under optimal policyholder behavior - results
found in the literature so far are consistent with the absence of
such a feature.  Recently, \citet{Azimzadeh2014} prove the existence
of an optimal \emph{bang-bang} control for a Guaranteed Lifelong Withdrawal
Benefits (GLWB) contract. In particular, they find that the holder
of a GLWB can maximize a writer�s losses by only ever performing
non-withdrawal, withdrawal at exactly the contract rate, or full
surrender. This dramatically reduces the optimal strategy space.
However, they also demonstrate that the related GMWB contract is not
convexity preserving, and hence does not satisfy the bang-bang
principle other than in certain degenerate cases.  For GMWB under
optimal withdrawal assumption, the numerical algorithms developed by
\citet{dai2008guaranteed}, \citet{Forsyth2008} and
\citet{LuoShevchenkoGMWB2015} appear to be the only ones found in
the literature, but none of them actually performed calculations
with surrender option on top of optimal withdrawal strategy. Also,
it is of practical interest to see how the much simpler bang-bang
strategy, although not optimal for GMWB, compares with optimal GMWB
strategy with surrender option.

Recently, in \citet{LuoShevchenkoGMWB2015}, we
have developed a new efficient numerical algorithm for pricing
GMWB contracts  in the case when transition density of the
underlying asset between withdrawal dates or its moments are known.
This algorithm  relies on computing the expected contract value
through a high order Gauss-Hermite
 quadrature applied on a cubic spline interpolation and much faster than the standard partial differential equation methods.  In this paper we extend our  algorithm to include surrender option in GMWB  and compare prices under
   different policyholder strategies: optimal, static and bang-bang.
   Results indicate that following a simple but sub-optimal bang-bang strategy does
 not lead to significant reduction  in the price or equivalently in the fee, in comparison with the optimal strategy.
 We also observed that the extra value added by the surrender option strongly depends on
volatility and the penalty charge,  among other factors such as
contractual rate, maturity and interest rate etc. At high volatility
or at low penalty
  charge, the surrender feature adds very significant value to the GMWB contract - the required fair fee is more than
  doubled in some cases; thus it is critical to account for surrender feature in pricing of real products. We also performed calculations for static withdrawal with surrender
option, which is the same as bang-bang minus the ``no-withdrawal"
 choice. We find that the fee for such contract is only less than $1\%$ smaller when compared to the case of bang-bang strategy,
 meaning that the
 ``no-withdrawal" option adds little value to the contract.

\end{abstract} 

\maketitle

\section{INTRODUCTION}
\label{sec:introduction} The world population is becoming older
fast. As a
consequence the age-related spending is projected to rise
dramatically in the coming decades in all the developed countries.
Increasingly governments in the developed world realize they cannot
afford paying sufficient public pensions and are looking for
innovations in retirement income product market.
 In this paper we consider a variable annuity contract with
Guaranteed Minimum Withdrawal Benefit (GMWB) with option to surrender the contract before maturity. This contract promises to return the
entire initial investment through cash withdrawals during the policy
life plus the remaining account balance at maturity, regardless of
the portfolio performance. Thus even  when the account of the
policyholder falls to zero before maturity, GMWB feature will
continue to provide the guaranteed cashflows. In addition, we allow the option to surrender the contract before the maturity which is a standard feature of real products on the market. GMWB allows the policyholder to withdraw funds below or at contractual rate without
penalty and above the contractual rate with some penalty. If the
policyholder behaves passively and the withdraw amount at each
withdrawal date
 is predetermined at
the beginning of the contract, then the behavior of the
 policyholder is  called ``static''. In this case the paths of the account  can be simulated  and a standard Monte Carlo
 simulation method can be used to price the GMWB. On the other hand if the policyholder optimally decide the amount of withdraw
 at each withdrawal date, then   the behavior of the policyholder is  called ``dynamic''.
Under the optimal withdrawal
strategy of a policyholder, the pricing of variable annuities with
GMWB becomes an optimal stochastic control problem; and adding surrender feature makes it also an optimal stopping problem.

The variable annuities with GMWB feature under dynamic and static withdrawal strategies have been considered in a
number of papers over the last decade, e.g. \citet{milevsky2006financial}, \citet{bauer2008universal}, \citet{dai2008guaranteed},
 \citet{Huang2012, Huang2014}, \citet{bacinello2011unifying}.
 Recently, \citet{Azimzadeh2014} prove the existence of an optimal \emph{bang-bang} control
for a Guaranteed Lifelong Withdrawal Benefits (GLWB) contract. In
particular, they find that the holder of a GLWB can maximize a
contract writer loss by only ever performing non-withdrawal, withdrawal
at exactly the contract rate, or full surrender. This dramatically
reduces the optimal strategy space. However, they also demonstrate
that the related GMWB contract is not convexity preserving, and
hence does not satisfy the bang-bang principle other than in certain
degenerate cases. For GMWB under optimal withdrawal
assumption, the numerical algorithms developed by
\citet{dai2008guaranteed} and \citet{Forsyth2008} appear to be the
only ones found in the literature, and both are based on solving
corresponding partial differential equation (PDE) via finite difference method.

 In the case when transition density of the underlying wealth process between
 withdrawal dates or its moments are known in closed form, often it can be
  more convenient and more efficient to utilize direct integration methods to calculate the
   required annuity value expectations in backward time-stepping procedure. Such an algorithm was developed in \citet{LuoShevchenkoGHQC2014, LuoShevchenkoGMWB2015} for solving optimal stochastic control problem in pricing GMWB variable annuity. This allows to get virtually instant results for typical GMWB annuity prices on the standard desktop PC. In this paper we adopt this algorithm to price
 variable  annuities with GMWB with surrender option under static, dynamic, and simplified bang-bang withdrawal strategies. To our best knowledge, there are no publications presenting results for GMWB with both optimal withdrawal and surrender features.

In the next section we describe the GMWB product with discrete
withdrawals, the underlying stochastic model and the optimization
problem. Section \ref{algorithm_sec} describes numerical algorithm utilized for pricing
 In Section \ref{NumericalResults_sec},  numerical results  for the fair fees
 under a series GMWB contract conditions are presented. Concluding remarks are given in Section \ref{conclusion_sec}.

\section{ Model}\label{model_sec}
We assume that market is complete in financial
risk and also there is no mortality risk (in the event of policyholder death, the contract is maintained by
beneficiary), thus the annuity price can be expressed as expectation under the risk neutral process of underlying asset. Let $S(t)$ denote the value of the reference portfolio of assets
(mutual fund index, etc.) underlying the variable annuity policy at
time $t$ that under no-arbitrage condition follows the risk neutral stochastic process
\begin{equation}\label{referenceportfolio_eq}
dS(t)=r(t) S(t) dt+\sigma(t) S(t) dB(t),
\end{equation}
where $B(t)$ is the standard Wiener process, $r(t)$ is risk free
interest rate and $\sigma(t)$ is volatility. For simplicity hereafter we assume that model parameters are piece-wise constant functions of time for time discretization $0=t_0<t_1<\cdots<t_N=T$, where $t_0=0$ is today and $T$ is annuity contract maturity. Denote corresponding asset values as $S(t_0),\ldots,S(t_N)$; and risk-free interest rate and volatility as $r_1,\ldots,r_N$ and $\sigma_1,\ldots,\sigma_N$ respectively. That is, $r_1$ is the interest rate for time teriod $(t_0,t_1]$; $r_2$ is for $(t_1;t_2]$, etc and similar for volatility.

The premium paid by policyholder upfront at $t_0$ is invested into the reference portfolio of risky assets $S(t)$.
Denote the value of this variable annuity account (hereafter
referred to as \emph{wealth account}) at time $t$ as $W(t)$, i.e.
the upfront premium paid by policyholder is $W(0)$. GMWB guarantees
the return of the premium via withdrawals $\gamma_n\ge 0$ allowed at
times $t_n$, $n=1,\ldots,N$.  Let $N_w$ denote the number of
withdrawals in a year (e.g. $N_w=12$ for a monthly withdrawal), then
the total number of withdrawals $N=\lceil\; N_w\times T \;\rceil$. The total of withdrawals cannot exceed the guarantee $W(0)$
and withdrawals can be different from contractual (guaranteed)
withdrawal $G_n=W(0)(t_n-t_{n-1})/T$,  with penalties imposed if
$\gamma_n>G_n$. Denote the annual contractual rate as $g=1/T$.

 Denote the value of the guarantee at time $t$ as $A(t)$, hereafter referred to as \emph{guarantee account}. Obviously, $A(0)=W(0)$. For clarity of notation, denote the time immediately before $t$ (i.e. before withdrawal) as  $t^-$, and  immediately
after $t$ (i.e. after withdrawal) as  $t^+$.
Then the guarantee balance
evolves as
\begin{equation}\label{accountbalance_eq}
A(t_n^+)=A(t_n^{-})-\gamma_n=A(t^+_{n-1})-\gamma_n,\;\; n=1,2,\ldots,N
\end{equation}
with $A(T^+)=0$, i.e. $W(0)=A(0) \ge \gamma_1+\cdots+\gamma_N$ and
$A(t_{n-1}^{+})\ge \sum_{k=n}^N\gamma_{k}$. The account balance $A(t)$ remains
unchanged within the interval $(t_{n-1},\;t_n), \;n=1,2,\ldots,N$.

In the case of reference portfolio process (\ref{referenceportfolio_eq}), the wealth account $W(t)$ evolves as
\begin{eqnarray}\label{eq_Wt}
W(t_n^-)&=&\frac{W(t_{n-1}^+)}{S(t_{n-1})}S(t_n) e^{-\alpha dt_n}=   W(t_{n-1}^+)e^{(r_n-\alpha-\frac{1}{2}\sigma^2_n)dt_n+\sigma_n \sqrt{dt_n} z_n},\\
W(t_n^+)&=&\max\left(W(t_n^-)-\gamma_n,0\right),\;\; n=1,2,\ldots,N,
\end{eqnarray}

where $dt_n=t_n-t_{n-1}$, $z_n$ are iid standard Normal random
variables and $\alpha$ is the annual fee charged by the insurance company. If the account balance
becomes zero or negative, then it will stay zero till maturity.


The cashflow received by the policyholder at withdrawal time $t_n$ is given by
\begin{equation}
C_n(\gamma_n)=\left\{\begin{array}{ll}
                   \gamma_n, & \mbox{if}\; 0\le \gamma_n\le G_n, \\
                   G_n+(1-\beta)(\gamma_n-G_n), & \mbox{if}\; \gamma_n>G_n,
                 \end{array} \right.
\end{equation}
where $G_n$ is contractual  withdrawal. That is, penalty is applied if
withdrawal $\gamma_n$ exceeds $G_n$, i.e. $\beta\in
[0,1]$ is the penalty applied to the portion of withdrawal above
$G_n$.

If the policyholder decides to surrender at time slice $\tau\in(1,\ldots,{N-1})$, then policyholder receives cashflow  $D_\tau(W(t_\tau),A(t_\tau))$ and contract stops. For numerical example we assume that
\begin{equation}\label{surrendercashflow_eq}
D_\tau(W(t_\tau),A(t_\tau)):=C_\tau(\max(W(t_\tau),A(t_\tau)));
\end{equation}
other standard surrender conditions can easily be implemented.

Denote the value of variable annuity at time $t$ as $Q_t(W(t),A(t))$, i.e. it is determined by values of the
wealth and guarantee accounts $W(t)$ and $A(t)$.  At maturity, if not surrendered earlier, the policyholder takes the maximum between
the remaining guarantee withdrawal net
of penalty charge and the remaining balance of the personal account,
i.e. the final payoff is
\begin{equation}
Q_{t_N^-}(W(T^-),A(T^-)):=h_N(W(T^-),A(T^-))=\max\left(W(T^-),C_N(A(T^-))\right).
\end{equation}

Under the above assumptions/conditions, the fair no-arbitrage value of the annuity at time $t_0$
is
\begin{eqnarray}\label{GMWB_general_eq}
&&\hspace{-1cm}Q_{t_0}\left ( W(t_0),A(t_0)\right)=\max_{\tau,\gamma_{1},\ldots,\gamma_{\widetilde{N}}}\mathrm{E}_{t_0}\bigg[B(0,\tau)D_\tau(W(t_\tau^-),A(t_\tau^-))\mathbb{I}_{\{t_\tau<T\}}\nonumber\\
&&\hspace{0cm}+B(0,N)h_N(W(T^-),A(T^-))(1-\mathbb{I}_{\{t_\tau<T\}})+\sum_{j=1}^{\widetilde{N}}
B(0,j)C_j(\gamma_j)\bigg], \;\;\widetilde{N}=\min(\tau,N)-1,
\end{eqnarray}
where $B_{0,n}=\exp(-\int_{0}^{t_n} r(\tau)d\tau)$ is discounting factor and $\mathbb{I}_{\{\cdot\}}$ is indicator function. Note that the today's value of the annuity policy $Q_0(W(0),A(0))$ is a function of policy fee
 $\alpha$.
Here, $\tau$ is stopping time and $\gamma_{1},\ldots,\gamma_{N-1}$ are the control variables chosen to maximize the expected value of discounted cashflows, and expectation $\mathrm{E}_0[\cdot]$ is taken under the risk-neutral process conditional on $W_0$ and $A_0$. The fair fee value of $\alpha$ corresponds to $Q_0\left(W(0),A(0)\right)=W(0)$. It is important to note that control variables and stopping time can be different for different realizations of underlying process and moreover the control variable $\gamma_n$ affects the transition law of the underlying wealth process from $t_n$ to $t_{n+1}$. Overall, evaluating GMWB with surrender feature is solving optimal stochastic control problem with optimal stopping.



Denote the state vector at time $t_n$ as $X_n=(W(t_n^-),A(t_n^-))$. Given that $\bm{X}=(X_1,\ldots,X_N)$ is Markov process, it is easy to recognize that the annuity valuation under the optimal withdrawal strategy (\ref{GMWB_general_eq}) is optimal stochastic control problem for Markov process that can be solved  recursively to find annuity value $Q_{t_n}(x)$ at $t_n$, $n=N-1,\ldots,0$ via backward induction
\begin{equation}
Q_{t_n}(x)=\max\left(\sup_{0\le\gamma_n\le A(t_{n}^-)}\left(C_n(\gamma_n(X_n))+ e^{-r_{n+1}dt_{n+1}}\int Q_{t_{n+1}}(x^\prime)K_{t_n}(dx^\prime|x,\gamma_n) \right),D_n(x)\right)
\end{equation}
starting from final condition $Q_T(x)=\max\left(W(T^-),C_N(A(T^-))\right)$. Here $K_{t_n}(dx^\prime|x,\gamma_n)$ is  the stochastic kernel representing probability to reach state in $dx^\prime$ at time $t_{n+1}$ if the withdrawal (\emph{action}) $\gamma_n$ is applied in the state $x$ at time $t_n$. For a good textbook treatment of stochastic control problem in finance, see \cite{bauerle2011markov}. Explicitly, this backward recursion can be solved as follows.


The annuity price at any time $t$ for a fixed $A(t)$  is a
function of $W$ only. Note $ A(t_{n-1}^+)= A(t_{n}^-)=A$ is constant in the period
$(t_{n-1}^+,t_n^-)$. Thus in a backward time-stepping setting (similar to a
finite difference scheme) the annuity value at
 time $t=t_{n-1}^+$  can be evaluated as the following expectation
\begin{equation}\label{eq_expS}
Q_{t^+_{n-1}}\left(W(t_{n-1}^+), A\right)=\mathrm{E}_{t_{n-1}}\left[e^{-r_n dt_n}
Q_{t_n^{-}}\left(W(t_n^-),A\right)|W(t_{n-1}^+),A\right].
\end{equation}
Assuming the conditional probability
distribution density of $W(t_n^-)$ given $W(t_{n-1}^+)$  is known as
$p_n(w(t_n)|w(t_{n-1}))$, then the above expectation can be evaluated
by
\begin{equation}\label{eq_intS}
Q_{t_{n-1}^+}\left(W(t_{n-1}^+), A\right)=\int_0^{+\infty}
e^{-r_n dt_n} p_n(w|W(t_{n-1}^+)) Q_{t_n^-}(w,A)dw.
\end{equation}
In the case of wealth process (\ref{eq_Wt}) the transition density $p_n(w(t_n)|w(t_{n-1}))$ is known in closed form and we will use Gauss-Hermite quadrature for the evaluation of the
above integration over an infinite domain. The required continuous
function $Q_t(W,A)$ will be approximated by a cubic spline
interpolation on a discretized grid in the  $W$ space.

Any change of $A(t)$ only occurs at withdrawal dates. After the
amount $\gamma_n$ is drawn at $t_n$, the wealth account reduces
from $W(t_n^-)$ to $W(t^+_n) = \max (W(t_n^-) -\gamma_n,0)$, and the
guarantee balance drops from $A(t_n^-)$ to $A(t_n^+)=A(t_n^-) -
\gamma_n$.
 Thus the jump condition of $Q_t(W,A)$ across $t_{n}$ is given
by
\begin{eqnarray}\label{eqn_jump}
&&\hspace{-1cm}Q_{t_{n}^-}(W(t_{n}^-),A(t_{n}^-))\nonumber\\
&&\hspace{-1cm}=\max\left(\max_{0 \leq \gamma_n\leq A(t_{n}^-) } [Q_{t_n^+}(\max(W(t_{n}^-)-\gamma_n,0),
A(t_{n}^-)-\gamma_n)+C_n(\gamma_n)],D_n(W(t_{n}^-),A(t_{n}^-))\right).
\end{eqnarray}
For optimal strategy, we chose a value for $\gamma_n$ under the
restriction $0 \leq \gamma_n\leq A(t_n^-) $ to maximize the function value
$Q_{t_n^-}(W,A)$ in (\ref{eqn_jump}). Repeatedly applying (\ref{eq_intS}) and (\ref{eqn_jump}) backwards in time starting from
\begin{equation}
Q_{t^-_N}(W(T^-),A(T^-))=\max\left(W(T^-),C_N(A(T^-))\right)
\end{equation}
gives us annuity value at $t=0$.

In additional to dynamic and static strategies, in this paper we also consider \emph{bang-bang} strategy which is simplified sup-optimal strategy where the policy holder at each $t_n$ can either make withdrawal at contractual rate $G_n$, make no withdrawal or surrender.

\section{Numerical algorithm}\label{algorithm_sec}
A very detailed description of the algorithm that we adapt for
pricing GMWB with surrender can be found in
\citet{LuoShevchenkoGMWB2015}. Below we outline the main steps.
We discretize the asset
domain $[W_{\min}, W_{\max}] $  by $W_{\min} =W_0 < W_1,
\ldots,W_M=W_{\max}$ , where $W_{\min}$ and $W_{\max}$ are the lower
and upper boundary, respectively.   The idea is to find annuity
values at all these grid points at  each time step from  $t_n^-$ to
$t_{n-1}^+$  through integration  (\ref{eq_intS}), starting at
maturity $t=t_N^-=T^-$.  At each time step we evaluate the
integration (\ref{eq_intS}) for every grid point  by a high accuracy
Gauss-Hermite numerical quadrature; it can also be accomplished by solving corresponding PDE using finite difference method that we implemented for benchmarking. At time step $t_n^- \rightarrow
t_{n-1}^+$, the annuity value at $t=t_n^-$ is known only at grid
points $W_m$, $m=0,1,\ldots,M$. In order to approximate the
continuous function $Q_t(W,A)$ from the values at the discrete grid
points, we
 use the cubic spline interpolation
 which is smooth in the first derivative and continuous in the second derivative.

For guarantee account balance variable $A$, we introduce an
auxiliary finite grid $0 = A_1 < \cdots < A_J = W(0)$ to track the
remaining guarantee balance  $A$, where $J$ is the total number of
nodes in the guarantee balance amount coordinate.  For each $A_j $,
we associate a continuous solution $Q_t(W,A_j)$.
 At every jump we let $A$ to be one of
the grid points $A_j ,\;1 \le j \le J$. For any $W=W_m$, $
m=0,1,\ldots, M$ and $A=A_j$, $ j=1,\ldots, J$ , given that
withdrawal amount can only take the pre-defined values
$\gamma=A_j-A_k$, $k=1,2,\ldots,j$, irrespective of time $t_n$ and
account value $W_m$, the jump condition (\ref{eqn_jump}) takes the
following form for the specific numerical setting
\begin{equation}\label{eqn_jump2}
Q_{t_n^-}(W_m,A_j)=\max\left(\max_{1\leq k \leq j}
[Q_{t_n^+}(\max(W_m-A_j+A_k,0), A_k)+C_n(A_j-A_k)],D_n(W_m,A_j)\right).
\end{equation}
Overall we have $J$  numerical solutions (obtained
through integration)  to track, corresponding to each of the $A_j$
value, $1\leq j \leq J$. Stepping backward in time, we find $Q_0(W(0),A(0))$ that depends on the policy fee $\alpha$. Finally, we calculate fair fee value of $\alpha$ corresponding to $Q_0(W(0),A(0))=W(0)$ that obviously requires iterative process.

\section{Numerical  Results}\label{NumericalResults_sec}
Below we present numerical results for fair fee of GMWB with surrender
option under  optimal and suboptimal bang-bang withdrawal
strategies. For convenience we denote results for optimal withdrawal
strategy without surrender option as GMWB, and with surrender option
as GMWB-S.
As discussed in \citet{LuoShevchenkoGMWB2015}, only very few results
for GMWB under dynamic policyholder behavior can be found in the
literature, and these results are for GMWB {\it without} the
surrender option. For validation purposes, perhaps the most accurate
results are  found in \citet{Forsyth2008}, which were obtained with
a very fine mesh in a detailed convergence study. As shown in Table
\ref{tab_g10}, our GMWB
  results for fair fee compare very well with those of  \citet{Forsyth2008}.  The maximum absolute
  difference in the fair fee rate between the two numerical studies is only $0.3$ basis point (a basis point is $0.01\%$ of a rate).

Table \ref{tab_g10} shows some very interesting comparison among
GMWB, GMWB-S and bang-bang results. At volatility $\sigma=0.2$, the
fair fee for GMWB-S  is virtually the same as GMWB, meaning
surrender adds little value to the optimal strategy; at high
volatility $\sigma=0.3$, fees for GMWB-S are significantly higher
than GMWB, up to $50\%$ higher at the half-yearly frequency. This
may suggest that at high volatility it is optimal to surrender at
high values of account balance or guarantee level. In addition, it
also shows higher frequency gives higher extra value to the
surrender option. Comparing bang bang with GMWB-S, the fees are
below the optimal strategy as expected, but the values are not very
significantly lower at both volatility values - it is only about
$10\%$ lower at most.

\begin{figure}[!htbp]
\begin{center}
\includegraphics[scale=0.55]{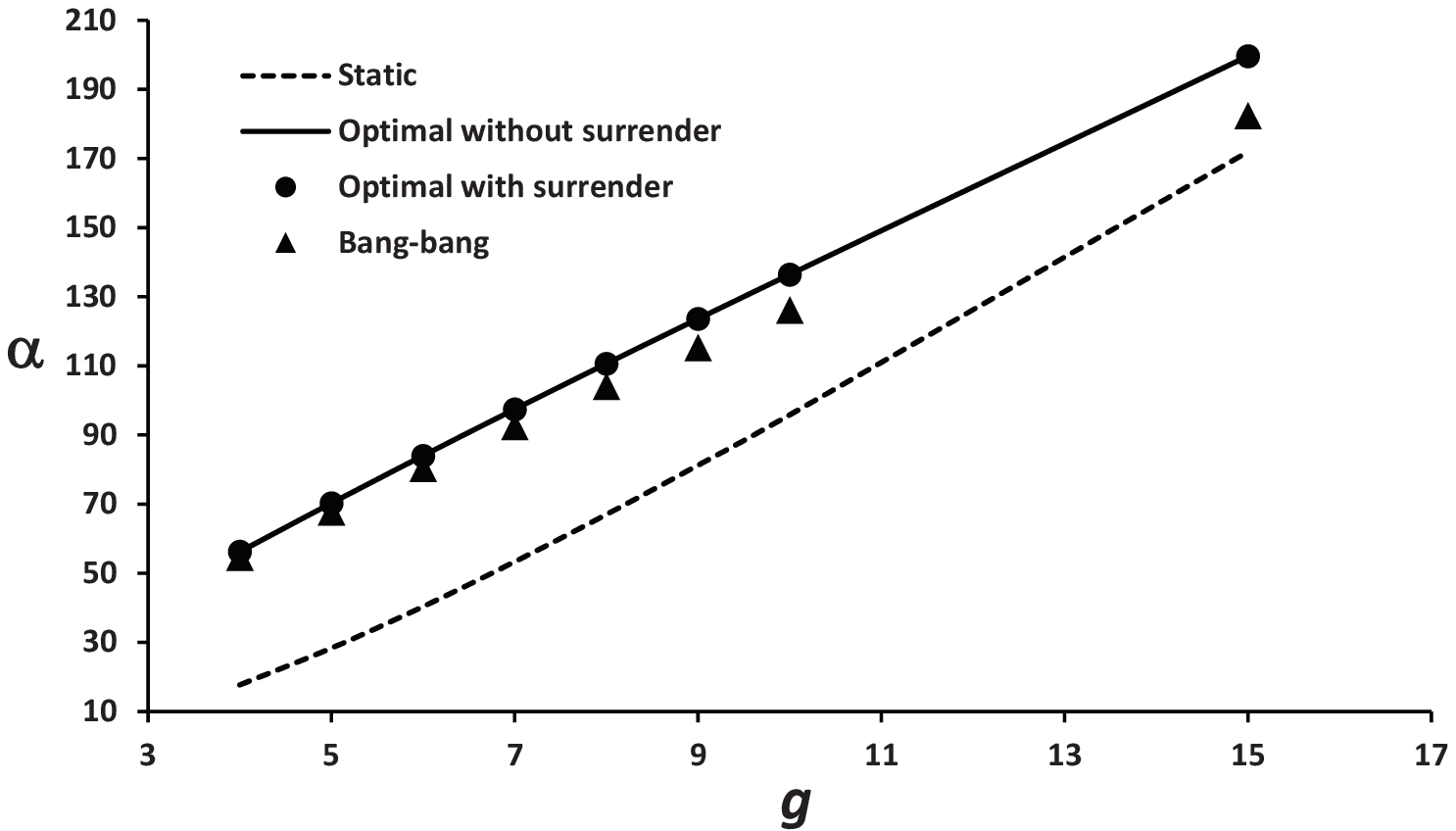}
\vspace{0cm} \caption{Fair fee $\alpha$ as a function of annual
guarantee rate $g$ for static, GMWB, GMWB-S and bang-bang at a
quarterly withdraw rate.  The fixed input parameters are
$\beta=10\%$, $r=5\%$ and $\sigma=0.2$.}\label{fig_fee1}
\end{center}
\end{figure}

\begin{figure}[!htbp]
\begin{center}
\includegraphics[scale=0.55]{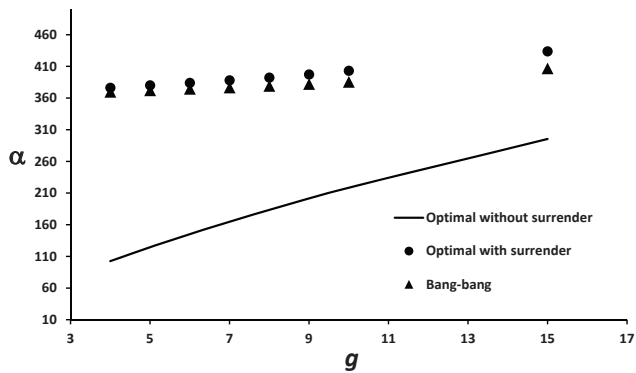}
\vspace{0cm} \caption{Fair fee $\alpha$ as a function of annual
guarantee rate $g$ for static, GMWB, GMWB-S and bang-bang at a
quarterly withdraw rate. The fixed input parameters are $\beta=5\%$,
$r=5\%$ and $\sigma=0.2$.}\label{fig_fee2}
\end{center}
\end{figure}

Figure \ref{fig_fee1} shows curves of the fee as a function of the
contractual annual withdraw rate, given $\sigma=0.2$, $r=0.05$ and
$\beta=0.1$. It compares  four cases: static (without surrender),
GMWB, GMWB-S and bang-bang, all at  quarterly withdrawal frequency
with $10\%$ penalty
    charge, i.e. $\beta=0.1$. This comparison  also shows GMWB-S and GMWB have virtually the same fees at $\sigma=0.2$, and
    bang-bnag is only slightly below GMWB-S, confirming results in Table
    \ref{tab_g10}. However, at the same volatility $\sigma=0.2$, new
    features emerge if we reduce the penalty
    charge from  $\beta=0.1$ to $\beta=0.05$, as shown in Figure
    \ref{fig_fee2}. When the penalty charge is reduced and all other parameters are unchanged, the
    surrender option adds more significant value to GMWB - in fact the fees are
     more than doubled at low to moderate contractual withdraw rate
     (or equivalently long or moderate maturity), i.e. fees for GMWB-S are more
     than twice as those for GMWB. With reduced penalty, fees for
     bang-bang are still close to the optimal strategy with
     surrender option, the GMWB-S.

We also performed calculations for static withdrawal with surrender
option, which is the same as bang-bang minus the ``no-withdrawal"
 choice. We find the fee for such contract is only less than $1\%$ smaller than the bang-bang strategy, meaning the
 ``no-withdrawal" option adds little value to the contract.
  Finally, different penalty functions can be applied to the
surrender (i.e. surrender cashflow can be different from (\ref{surrendercashflow_eq})). For example, instead of penalizing only the amount
exceeding the contractual withdrawal rate, we can penalize the
entire termination amount. In this case we find both GMWB-S and
bang-bang yield only slightly lower fees for a given $\beta$ - this
is perhaps  not very surprising since when it is optimal to
surrender, the amount must be much higher than the contractual rate,
thus penalizing the entire amount is not much more severe than
penalizing only the exceeded part.

\begin{table}[!htbp]
\begin{center}
{{\begin{tabular*}{0.75\textwidth}{cccccc} \hline
 frequency & volatility &  Chen \& Forsyth &   GMWB  & GMWB-S & Bang Bang \\
 \hline
 yearly      & 0.2 & 129.1 & 129.1  & 129.2 & 123.9  \\
 half-yearly & 0.2 & 133.5 & 133.7  & 134.0  & 125.6  \\
 yearly      & 0.3 & 293.3 & 293.5   & 418.4  & 392.9 \\
 half-yearly & 0.3 & 302.4 & 302.7   & 456.5  & 410.7   \\
 \hline
\end{tabular*}
}}\end{center} \caption{Comparison of fair fee $\alpha$ in basis points (a basis point is 0.01\%) between
results of  GMWB, GMWB-S and bang-bang. Results under  ``Chen \&
Forsyth" are for GMWB.  The input parameters are $g=10\%$,
$\beta=10\%$, $r=5\%$ and $\sigma=0.2$. The withdrawal frequency is
quarterly. } \label{tab_g10}
\end{table}

\newpage

\section{CONCLUSIONS}\label{conclusion_sec}
 In this
paper we have developed numerical valuation of variable
annuities with GMWB and surrender features under both static, dynamic (optimal) and bang-bang
policyholder strategies. The results indicate that following a simple bang-bang strategy does
 not lead to significant reduction  in the price or equivalently in the fee. We also observed that
the extra value added by the surrender option very much depends on
volatility and the penalty charge, among other facts such as
contractual rate and maturity.
 At high volatility or at
 low penalty
  charge, the surrender feature adds very significant value to the GMWB contract - more than doubling in some cases; highlighting the importance of accounting for surrender feature in pricing of real products.
We have assumed the policyholder will always live beyond the
maturity date or there is always someone there to make optimal
withdrawal decisions for the entire duration of the contract. It is not difficult to  consider
adding some death benefits on top of GMWB, i.e. combining GMWB with
some kind of life insurance as it is done in our recent paper
\citet{LuoShevchenkoGMWDB2015} considering both market
process and death process. Further work includes admitting other
stochastic risk factors such as stochastic interest rate or
volatility.

\section*{Acknowledgement}
We gratefully acknowledge financial support by the CSIRO-Monash
Superannuation Research Cluster, a collaboration among CSIRO, Monash University, Griffith University, the University of Western Australia, the University of Warwick, and stakeholders of the retirement system in the interest of better outcomes for all.

\bibliography{bibliography}

\begin{thebibliography}{}

\bibitem[\protect\citeauthoryear{Azimzadeh and Forsyth}{Azimzadeh and
  Forsyth}{2014}]{Azimzadeh2014}
Azimzadeh, Y. and P.~A. Forsyth (2014).
\newblock The existence of optimal bang-bang controls for gmxb contracts.
\newblock Working paper of University of Waterloo.

\bibitem[\protect\citeauthoryear{Bacinello, Millossovich, Olivieri, and
  Pitacco}{Bacinello et~al.}{2011}]{bacinello2011unifying}
Bacinello, A., P.~Millossovich, A.~Olivieri, and E.~Pitacco (2011).
\newblock Variable annuities: a unifying valuation approach.
\newblock {\em Insurance: Mathematics and Economics\/}~{\em 49\/}(1), 285--297.

\bibitem[\protect\citeauthoryear{Bauer, Kling, and Russ}{Bauer
  et~al.}{2008}]{bauer2008universal}
Bauer, D., A.~Kling, and J.~Russ (2008).
\newblock A universal pricing framework for guaranteed minimum benefits in
  variable annuities.
\newblock {\em ASTIN Bulletin\/}~{\em 38\/}(2), 621--651.

\bibitem[\protect\citeauthoryear{B\"{a}uerle and Rieder}{B\"{a}uerle and
  Rieder}{2011}]{bauerle2011markov}
B\"{a}uerle, N. and U.~Rieder (2011).
\newblock {\em Markov Decision Processes with Applications to Finance}.
\newblock Springer, Berlin.

\bibitem[\protect\citeauthoryear{Chen and Forsyth}{Chen and
  Forsyth}{2008}]{Forsyth2008}
Chen, Z. and P.~Forsyth (2008).
\newblock A numerical scheme for the impulse control formulation for pricing
  variable annuities with a guaranteed minimum withdrawal benefit (gmwb).
\newblock {\em Numerische Mathematik\/}~{\em 109\/}(4), 535--569.

\bibitem[\protect\citeauthoryear{Dai, Kwok, and Zong}{Dai
  et~al.}{2008}]{dai2008guaranteed}
Dai, M., K.~Y. Kwok, and J.~Zong (2008).
\newblock Guaranteed minimum withdrawal benefit in variable annuities.
\newblock {\em Mathematical Finance\/}~{\em 18\/}(4), 595--611.

\bibitem[\protect\citeauthoryear{Huang and Forsyth}{Huang and
  Forsyth}{2012}]{Huang2012}
Huang, Y. and P.~A. Forsyth (2012).
\newblock Analysis of a penalty method for pricing a guaranteed minimum
  withdrawal benefit ({GMWB}).
\newblock {\em Journal of Numerical Analysis\/}~{\em 32}, 320--351.

\bibitem[\protect\citeauthoryear{Huang and Kwok}{Huang and
  Kwok}{2014}]{Huang2014}
Huang, Y. and Y.~K. Kwok (2014).
\newblock Analysis of optimal dynamic withdrawal policies in withdrawal
  guarantee products.
\newblock {\em Journal of Economic Dynamics and Control\/}~{\em 45}, 19--43.

\bibitem[\protect\citeauthoryear{Luo and Shevchenko}{Luo and
  Shevchenko}{2014}]{LuoShevchenkoGHQC2014}
Luo, X. and P.~V. Shevchenko (2014).
\newblock Fast and simple method for pricing exotic options using gauss-hermite
  quadrature on a cubic spline interpolation.
\newblock {\em Journal of Financial Engineering\/}~{\em 1\/}(4).
\newblock DOI: 10.1142/S2345768614500330.

\bibitem[\protect\citeauthoryear{Luo and Shevchenko}{Luo and
  Shevchenko}{2015a}]{LuoShevchenkoGMWB2015}
Luo, X. and P.~V. Shevchenko (2015a).
\newblock Fast numerical method for pricing of variable annuities with
  guaranteed minimum withdrawal benefit under optimal withdrawal strategy.
\newblock {\em To appear in Journal of Financial Engineering, preprint is
  available at http://arxiv.org/abs/1410.8609\/}.

\bibitem[\protect\citeauthoryear{Luo and Shevchenko}{Luo and
  Shevchenko}{2015b}]{LuoShevchenkoGMWDB2015}
Luo, X. and P.~V. Shevchenko (2015b).
\newblock Valuation of variable annuities with guaranteed minimum withdrawal
  and death benefits via stochastic control optimization.
\newblock {\em Insurance: Mathematics and Economics\/}~{\em 62}, 5--15.

\bibitem[\protect\citeauthoryear{Milevsky and Salisbury}{Milevsky and
  Salisbury}{2006}]{milevsky2006financial}
Milevsky, M.~A. and T.~S. Salisbury (2006).
\newblock Financial valuation of guaranteed minimum withdrawal benefits.
\newblock {\em Insurance: Mathematics and Economics\/}~{\em 38\/}(1), 21--38.

\end{thebibliography}
\bibliographystyle{chicago} 
\end{document}